\documentclass[graybox]{svmult}

\usepackage{mathptmx}       
\usepackage{helvet}         
\usepackage{courier}        
\usepackage{type1cm}        
\usepackage{makeidx}         
\usepackage{graphicx}        
\usepackage{multicol}        
\usepackage[bottom]{footmisc}

\def\lesssim{\mathrel{\hbox{\rlap{\hbox{\lower4pt\hbox{$\sim$}}}\hbox{$<$}}}}
\def\gtrsim{\mathrel{\hbox{\rlap{\hbox{\lower4pt\hbox{$\sim$}}}\hbox{$>$}}}}

\makeindex             

\begin{document}

\title*{Disk-Magnetosphere Interaction and Outflows: Conical Winds
and Axial Jets}
\titlerunning{Disk-Magnetosphere Interaction and Outflows}
\author{M.M. Romanova, G.V. Ustyugova, A.V. Koldoba,
 \& R.V.E. Lovelace}
\institute{M.M. Romanova \at Dept. of Astronomy, Cornell University,
Ithaca, NY 14853 \email{romanova@astro.cornell.edu} \and G.V.
Ustyugova \at Keldysh Inst. of the Applied Math. RAS, Moscow,
125047, Russia \email{ustyugg@rambler.ru} \and A.V. Koldoba \at
Institute for Mathematical Modeling RAS, Moscow, 125047, Russia
\email{koldoba@rambler.ru} \and R.V.E. Lovelace \at Dept. of
Astronomy, Cornell University, Ithaca, NY 14853
\email{RVL1@cornell.edu}}

%
%
\maketitle

\abstract{We investigate outflows from the disk-magnetosphere
boundary of rotating magnetized stars in cases where the magnetic
field of a star is bunched into an X-type configuration using
axisymmetric and full 3D MHD simulations. Such configuration appears
if viscosity in the disk is larger than diffusivity, or if the
accretion rate in the disk is enhanced. Conical outflows flow from
the inner edge of the disk to a narrow shell with an opening angle
30-45 degrees. Outflows carry 0.1-0.3 of the disk mass and part of
the disk's angular momentum outward. Conical outflows appear around
stars of different periods, however in case of stars in the
``propeller" regime, an additional - much faster component appears:
an axial jet, where matter is accelerated up to very high velocities
at small distances from the star by magnetic pressure force above
the surface of the star. Exploratory 3D simulations show that
conical outflows are symmetric about rotational axis of the disk
even if magnetic dipole is significantly misaligned. Conical
outflows and axial jets may appear in different types of young stars
including Class I young stars, classical T Tauri stars, and EXors.}

\begin{figure}[b]
\centering
\includegraphics[scale=0.8]{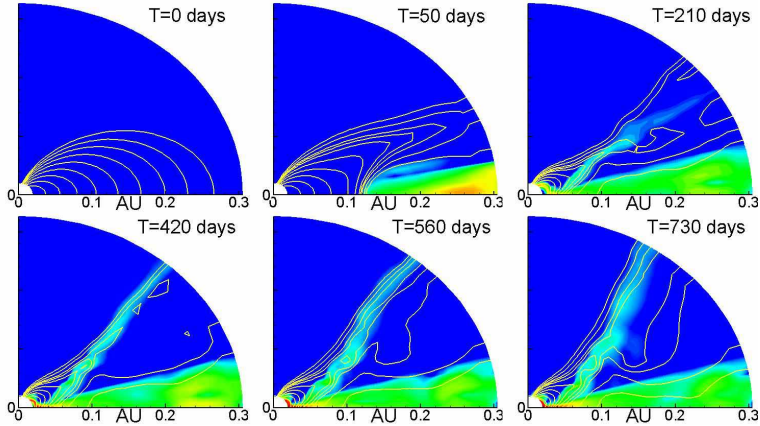}
\caption{Snapshots from axisymmetric simulations of conical winds.
The background shows the matter flux with light color corresponding
to higher flux. The lines are magnetic field lines.}
\label{dp-6}       
\end{figure}

\begin{figure}[t]
\centering
\includegraphics[scale=0.8]{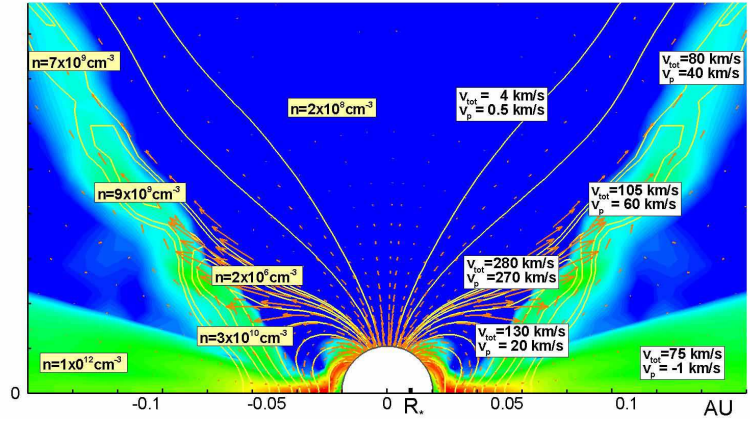}
\caption{Typical flow in conical winds (at $t=380$ days). The
background shows matter flux, lines are selected field lines, arrows
are proportional to velocity. The numbers show poloidal $v_p$ and
total $v_{tot}$ velocities and number density at sample places of
the simulation region.} \label{con-numb}
\end{figure}

\begin{figure}[b]
\centering
\includegraphics[scale=0.7]{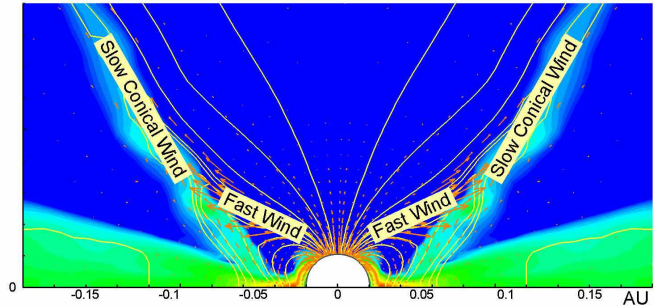}
\caption{Two components of winds from slowly rotating star are
labeled.}
\label{sym-label}       
\end{figure}

\begin{figure}[b]
\centering
\includegraphics[scale=0.2]{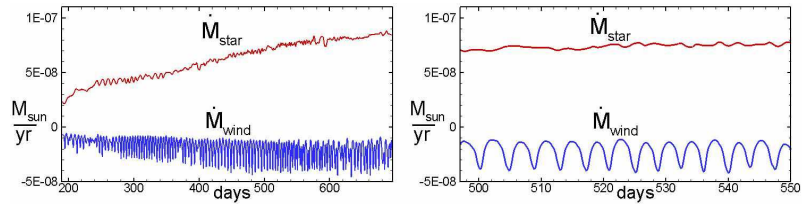}
\caption{Left panel: matter flux to the star $\dot M_{star}$ and
 to conical wind $\dot M_{wind}$ (calculated at the radius
$R=0.1$ AU) as function of time. Right panel: same but for a shorter
time-interval.} \label{flux-con}
\end{figure}

\section{Introduction}

 Jets\index{jets} and winds\index{winds} are observed in young stars\index{young stars}
 at different stages of their evolution from very young stars up to classical T Tauri
stars\index{classical T Tauri stars} (CTTSs) where
 smaller-scale jets and winds are observed
 (see review by Ray et al. 2007). A significant number of CTTS
 show signs of outflows\index{outflows} in spectral lines, in particular in
He I (Edwards et al. 2006; Kwan, Edwards, \& Fischer 2007).
High-resolution observations show that outflows often have an
``onion-skin" structure, with high-velocity outflows in the axial
region, and lower-velocity outflow at larger distance from the axis
(Bacciotti et al. 2000). High angular resolution spectra of
[FeII]$\lambda$ 1.644$\mu$m emission line taken along the jets from
DG Tau, HL Tau and RW Auriga revealed two components: a
high-velocity well-collimated extended component with v $\sim
200-400$ km/s and a low-velocity $\sim 100$ km/s uncollimated
component which is close to a star (Pyo et al. 2003, 2006).
High-resolution observations of molecular hydrogen in HL Tau have
shown that at small distances from the star the flow shows a conical
structure with outflow velocity $50-80$ km/s (Takami et al. 2007).

 Different models have been
proposed to explain outflows from CTTSs (see review by Ferreira,
Dougados, \& Cabrit 2006), including models where the outflow
originates from the inner regions of the accretion
disk\index{accretion disk} (e.g., Lovelace, Berk \& Contopoulos
1991; K\"onigl \& Pudritz 2000; Ferreira et al. 2006), and the
X-wind\index{x-wind} type models (Shu et al. 1994; 2007; Najita \&
Shu 1994; Cai et al. 2008) where most of the matter flows from the
disk-magnetosphere boundary\index{disk-magnetosphere boundary}. In
this work we consider only the second type of models. We developed
conditions favorable for X-type outflows and performed axisymmetric
and exploratory 3D MHD simulations\index{MHD simulations} for both
slowly and rapidly rotating stars including stars in the propeller
regime\index{propeller regime}.

\begin{figure}[t]
\centering
\includegraphics[scale=0.8]{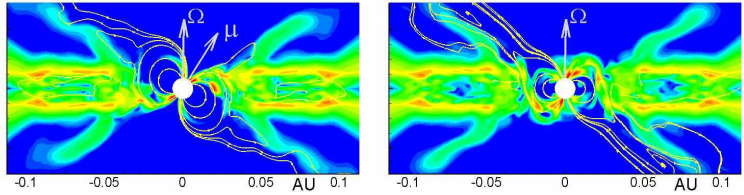}
\caption{Conical winds obtained in 3D MHD simulations for
$\Theta=30^\circ$. Left panel: density distribution and sample field
lines in the $\mu\Omega$ - plane. Right panel: same but in the
perpendicular plane.} \label{3d-wind}
\end{figure}

\section{Conical Winds}

\noindent{\bf Axisymmetric (2.5D) simulations}. To investigate
outflows from the disk-magnetosphere boundary it was important that
the magnetic field lines be bunched into an X-type configuration.
Such bunching will occur if magnetic field lines threading the disk
move inward to the star faster than they diffuse outward. This
happens for example when the viscosity in the disk is larger than
the diffusivity. In axisymmetric simulations we have both viscosity
and diffusivity incorporated in the code, both in $\alpha$ -
prescription (Shakura \& Sunyaev 1973). The coefficients $\alpha_v$
and $\alpha_d$ control these processes (Romanova, et al. 2005;
Ustyugova et al. 2006). We investigate a wide range of parameters:
$0.01<\alpha_v < 1$ and $0.01<\alpha_d < 1$ and choose $\alpha_v =
0.03$ and $\alpha_d = 0.1$ as a main case. We assume that after
period of low accretion rate the disk matter comes to the region
from the boundary. Matter efficiently bunches field lines and in our
case $\alpha_v>\alpha_d$ this configuration exists for a long time.
The disk matter comes close to the star, is stopped by the
magnetosphere, and part of it moves into persistent conical
outflows\index{conical outflows} (see Fig. 1).

Our simulations are dimensionless. As an example we chose parameters
of the typical CTTS with mass $M_*=0.8 M_\odot$, $R_*=2 R_\odot$,
magnetic field $B_*=1 kG$, period $P_*=5.4$ days. In the Figs. 1-3
the inner boundary corresponds to two radii of the star. We accepted
this choice of units so as to compare results with the propeller
case (see $\S 4$) where the inner boundary is a factor of two
smaller.
 Analysis of conical winds done by Romanova
et al. (2009) have shown that they are driven mainly by the magnetic
pressure force (e.g., Lovelace et al. 1991) which is largest right
above the disk and acts up to distances of about 12 stellar radii.
Fig. 2 shows typical parameters in a conical wind. Fig. 2 shows that
matter start to flow to a conical wind with very high azimuthal
velocity, equal to Keplerian velocity at the base of the outflow
($v_\phi\approx 130$ km/s in our main case). The poloidal velocity
increases along the flow from few km/s right above the disk up to
$v_p=40-60$ km/s at larger distances. Azimuthal velocity remains
larger than poloidal velocity inside the simulation region. In the
conical wind matter flows into a relatively narrow shell and the
cone has an opening angle, $\theta=30^\circ-40^\circ$. This may be
explained by the fact that the magnetic pressure force acts almost
vertically. This may also explain frequent events of reconnection of
the inflating magnetic field lines in the outflow. We note that in
addition to the main conical wind there is matter acceleration along
magnetic field lines closer to the axis. The low-density matter is
accelerated up to hundreds of km/s right near the star and may be
important in explanation of some highly blue-shifted spectral lines
which form near CTTSs. Matter which is accelerated in this region
may come from the star, or may be partially captured from the main
accretion flow. Fig. 3 shows two components of the flow around a
slowly rotating star.
\smallskip

\noindent{\bf The fluxes of matter and angular momentum} flowing to
or out from the star and fluxes flowing with conical winds through
the surface with radius $R = 0.1$AU were calculated. Fig. 4 shows
that matter flux to the wind is only several times smaller than that
to the star, $\dot M_{wind}\approx 0.2-0.3 \dot M_{star}$. The
matter flux going to the wind varies, which is connected with
frequent events of reconnection of the magnetic flux. It is often
the case that matter is outbursted to the conical winds in an
oscillatory regime, in particular if $\alpha_v$ and $\alpha_d$ are
not very small, $\alpha_{v,d}\sim 0.1-0.3$. If the diffusivity is
small, $\alpha_d=0.01-0.03$, then outbursts to winds are sporadic
and occur with a longer time-scale. Analysis of the angular momentum
shows that in the case of a slowly rotating star the star spins-up
by accreting matter (through magnetic torque at the surface of the
star, e.g. Romanova et al. 2002). Conical winds carry away part of
the angular momentum of the disk (0.5 in this example), however a
star may spin-up or spin-down depending on $P_*$. It spins-up in our
example of a slowly rotating star. We also checked the case of very
slow rotation, $P_*=11$ days, and observed that persistent conical
winds form in this case as well.

\smallskip

\noindent{\bf 3D simulations.} We performed exploratory 3D MHD
simulations of conical winds in the case where the dipole magnetic
field is misaligned relative to rotational axis by an angle
$\Theta=30^\circ$. Compared with the axisymmetric simulations, the
accretion disk is situated at $r > 10 R_*$ and the simulation region
is much larger. Viscosity is incorporated in the code and we chose
$\alpha_{vis}=0.3$ while the diffusivity is not incorporated and is
only numerical (small, at the level $\alpha_d=0.01-0.02$ at the
disk-magnetosphere boundary). We observed that the disk moved
inward, bunched field lines and formed conical winds. Fig. 5 shows
that conical winds are approximately symmetric about rotation axis.
There is however enhancement in the density distribution inside
conical winds which is associated with a spiral wave generated by
the misaligned dipole. Recent 3D modeling have shown that at a wide
range of parameters matter penetrates through the magnetosphere due
to interchange instability (Romanova, Kulkarni \& Lovelace 2008;
Kulkarni \& Romanova 2008). 3D simulations of conical wind show that
formation of conical winds occurs at larger distances from the star
and are not influenced by instabilities.

\begin{figure}[t]
\centering
\includegraphics[scale=0.7]{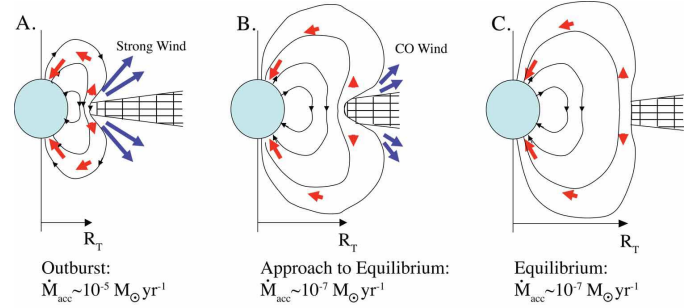}
\caption{Schematic model of an Exor V1647 Ori. During the outburst
the accretion rate is enhanced so that the magnetospheric radius
$R_m$ decreases and the magnetic field lines were bunched (A). This
results in a fast, hot outflow. As the accretion rate decreases, the
disk moves outward and this results in a slower, cooler CO outflow
(B). Further decrease in the accretion rate leads to a quiescence
state where the production of warm outflows stops (C). From Brittain
et al. (2007).} \label{brittain}
\end{figure}

\begin{figure}[t]
\centering
\includegraphics[scale=0.8]{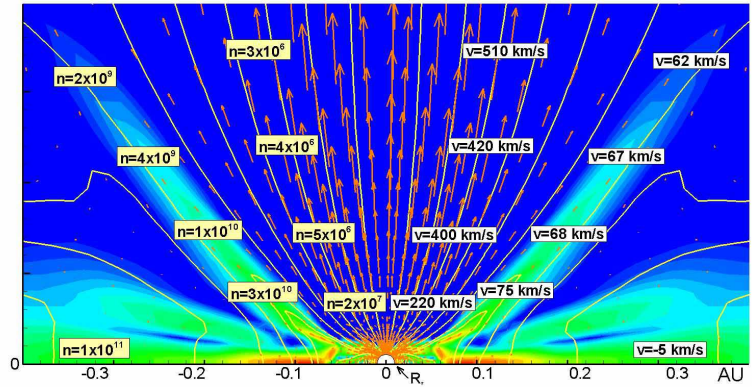}
\caption{Outflows in the propeller regime. The background shows
matter flux, lines are selected field lines, arrows are proportional
to velocity. Labels show total velocity and density at sample
points.} \label{prop-numb}
\end{figure}

\begin{figure}[b]
\centering
\includegraphics[scale=0.7]{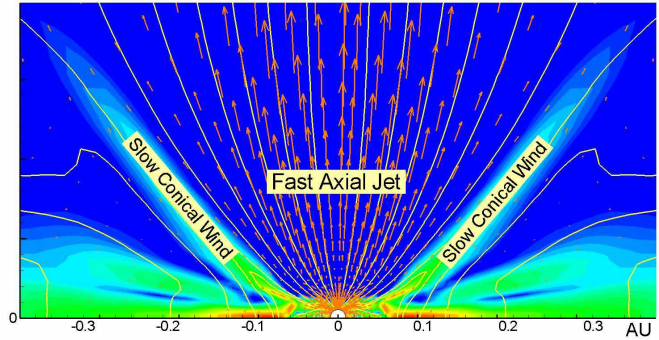}
\caption{Two components of outflows in the propeller regime.}
\label{prop-label}       
\end{figure}

\begin{figure}[b]
\centering
\includegraphics[scale=0.2]{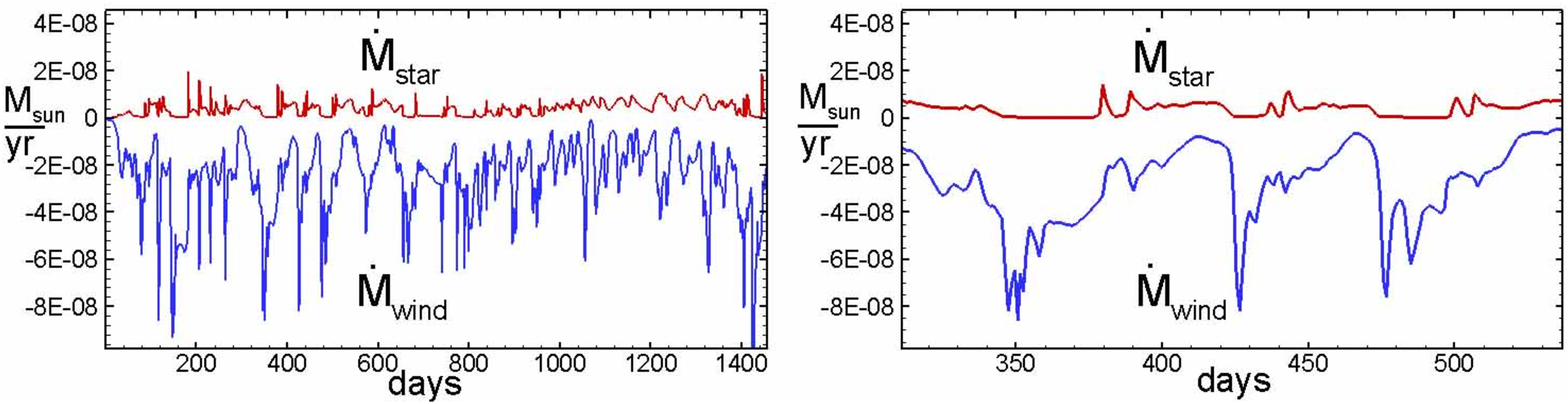}
\caption{Left panel: matter fluxes to the star $\dot M_{star}$ and
to the conical wind $\dot M_{wind}$ (calculated at $R=0.1$ AU) as
function of time. Right panel: same but for a shorter
time-interval.} \label{flux-prop}
\end{figure}

\section{\bf Enhanced Accretion and Outflows}

 CTTSs are strongly variable on different time-scales
including a multi-year scale (Herbst et al. 2004; Grankin et al.
2007). This is connected with variation of the accretion rate
through the disk which may lead to the enhancement of outflows
(e.g., Cabrit et al. 1990). Simulations have shown that the bunching
of field lines by the new matter after period of the low-density
accretion may lead to quite long outburst of matter to the conical
winds and may be the reason for formation of micro-jets in the
CTTSs. If CTTS is in a binary system, then an accretion rate may be
episodically enhanced due to interaction with the secondary star.
Events of fast, implosive accretion are possible due to thermal
instability or global magnetic instability, where the accretion rate
is enhanced due to the formation of disk winds (Lovelace, Romanova,
\& Newman 1994). Enhanced accretion may lead to outbursts in EXors,
where the accretion rate increases up to $10^{-5} M_{\odot}/yr$ and
strong outflows are observed. Brittain et al. (2007) reported on the
outflow of warm gas  from the inner disk around EXor V1647 observed
in the blue absorption of the CO line during the decline of the EXor
activity. He concluded that this outflow is a continuation of
activity associated with early enhanced accretion and bunching of
the field lines (see Fig. 6). In our main example of a CTTS the disk
stops at $R_m=2.4R_*$. In EXors, we take the radius of a star at the
Figs. 1-3 equal to the inner boundary, so that the disk stops much
closer to the star, $R_m=1.2 R_*$. Then all velocities are a factor
1.4 higher and densities a factor of 32 higher (compared to Figs. 2,
7), and matter fluxes in Figs. 4 and 9 are a factor of 11 higher
than in the main example relevant to CTTSs.

\section{Outflows in the ``Propeller" Regime}

In the propeller regime the magnetosphere rotates faster than inner
region of the disk. This occurs if the co-rotation radius
$R_{cr}=(GM/\Omega_*^2)^{1/3}$ is smaller than magnetospheric radius
$R_m$ (e.g., Lovelace et al. 1999). Young stars are expected to be
in the propeller regime in two situations: (1) At the early stages
of evolution (say, at $T < 10^6$ years), when the star formed but
did not have time to spin-down, and (2) at later stages of
evolution, such as at CTTS stage, when the star is expected to be on
average in the rotational equilibrium state (e.g., Long et al. 2005)
but variation of the accretion rate leads to variation of $R_m$
around $R_{cr}$, where $R_{cr} < R_m$ is possible. We performed
axisymmetric simulations of accretion to a star in the propeller
regime, taking a star with the same parameters as in case of conical
winds, but with period $P_*=1$ day (Romanova et al. 2005; Ustyugova
et al. 2006). We chose $\alpha_v=0.3$ and $\alpha_d=0.1$ and thus
bunched the field lines to the X-type configuration  We observed
that in addition to conical wind there is a fast axial jet (see Fig.
7) so that the outflow has two components (Fig. 8). The conical wind
in this case is much more powerful - it carries most of the disk
matter away. The axial jet carries less mass, but it is accelerated
to high velocities. Acceleration occurs due to the magnetic pressure
of the ``magnetic tower" which forms above the star as a result of
winding of magnetic field lines of the star. Outbursts to conical
winds occur sporadically with a long time-scale interval (see Fig.
9) which is connected with the long time-scale interval of
accumulation and diffusion of the disk matter through the
magnetosphere of the star (see also Goodson et al. 1997; Fendt
2008). These propeller outflows were obtained in conditions
favorable for such a process: when the star rotated fast and an
X-type configuration developed. Future simulations should be done
for the case of propeller-driven outflows from slower rotating CTTS.
Collimation of conical winds may occur at larger distances from the
star for example, by disk winds (e.g., K\"onigl \& Pudritz 2000;
Ferreira et al. 2006; Matsakos et al. 2008).

\section{Conclusions}

We discovered a new type of outflows - conical winds - in numerical
simulations where magnetic field lines are bunched into an X-type
configuration. In many respects these winds are similar to X-winds
proposed by Shu and collaborators (e.g., Shu et al. 1994): (1) They
both require bunching of the field lines; (2) They both have high
rotation of the order of Keplerian rotation at the base of outflow,
and gradual poloidal acceleration; (3) They both are driven by
magnetic force. However, there are a number of important
differences: (1) Conical winds flow in a thin shell, while X-winds
flow at different angles  below the ``dead zone"; (2) Conical winds
form around stars of any rotation rate including slow rotation, and
do not require the fine tuning of angular velocity of the inner disk
to that of magnetosphere; (3) Conical winds are non-stationary: the
magnetic field constantly inflates and reconnects; (4) Conical winds
carry away part of the angular momentum of the inner disk and are
not responsible for spinning-down the star, while X-winds are
predicted to take away angular momentum from the star and thus to
solve the angular momentum problem; (5) In conical winds there is a
fast component of the flow along field lines threading the star.
Some of these differences, such as non-stationarity of conical winds
is connected with natural restrictions of the stationary model of
X-winds. Conical winds can explain conical shape of outflows near
young stars of different type (CTTSs, EXors, Type I objects) which
have been recently resolved. In another example, Alencar et al.
(2005) analyzed blue-shifted absorption of  $H_\beta$ line in RW
Aurigae and concluded that conical shape wind with opening angle
$30-40^\circ$ and narrow annulus gives best match to the
observations of this line (see Fig. 10).

\begin{figure}[t]
\sidecaption
\includegraphics[scale=0.35]{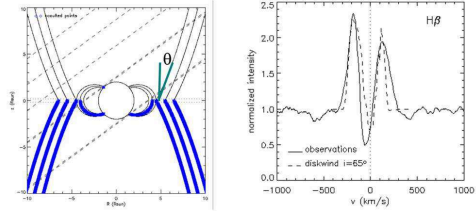}
\caption{Modeling of the  $H_\beta$ line in RW Aurigae led to the
conclusion that a conical shaped wind with opening angle
$30-40^\circ$ and a narrow annulus gives the best match to the
observations of this line (from Alencar et al. 2005).}
\label{rw_aur}       
\end{figure}

In the propeller regime the flow has two components: (1) a rapidly
rotating, relatively slow, dense conical wind, and (2) a fast, lower
density axial jet where matter is accelerated by magnetic pressure
up to hundreds of km/s very close to the star. Young stars of
classes 0 and I may be in the propeller regime and can lose most of
their angular momentum by this mechanism (Romanova et al. 2005). Or
any slower rotating magnetized stars  may enter the propeller regime
if the accretion rate becomes sufficiently low and the
magnetospheric radius becomes larger than the corotation radius. The
last possibility requires additional numerical simulations and
analysis.

\begin{acknowledgement}
The authors were supported in part by NASA grant NNX08AH25G and by
NSF grants AST-0607135 and AST-0807129. MMR thanks NASA for use of
the NASA High Performance Facilities. AVK and GVU were supported in
part by grant RFBR 06-02016608, Program 4 of RAS. MMR and RVEL thank
the organizers for a very interesting meeting and MMR is grateful to
the organizers for the generous support.
\end{acknowledgement}

\end{document}